\documentclass[aip,jmp,amsmath,amssymb]{revtex4-1}
\usepackage{graphicx}

\draft 

\begin{document}

\newcommand\ot{\otimes}
\newcommand\ttt{{\text{\rm t}}}
\newcommand\tr{{\text{\rm Tr}}\,}
\newcommand\meet{\wedge}
\newcommand\lan{\langle}
\newcommand\ran{\rangle}
\newcommand\la{\lambda}
\newcommand\wt{\widetilde}

\title{The structural physical approximations and optimal entanglement witnesses} 

\author{Kil-Chan Ha}

\affiliation{Faculty of Mathematics and Statistics,\\ Sejong University, Seoul 143-747, Korea}
\author{Seung-Hyeok Kye}
\affiliation{Department of Mathematics and Institute of Mathematics, \\Seoul National University, Seoul 151-742, Korea}


\date{\today}

\begin{abstract}
We introduce the notions of positive and copositive types for
entanglement witnesses, depending on the distance to the positive
part and copositive part. An entanglement witness $W$ is of positive
type if and only if its partial transpose $W^\Gamma$ is of
copositive type. We show that if the structural physical
approximation of $W$ is separable then $W$ should be of copositive
type, and the SPA of $W^\Gamma$ is never separable unless $W$ is of
both positive and copositive type. This shows that the SPA
conjecture is meaningful only for those of copositive type. We
provide examples to show that the SPA conjecture fails even for the
case of copositive types.
\end{abstract}


\maketitle 

\section{Introduction}
Quantum entanglement is the key resource for applications to quantum
computation and quantum information theory. See Ref.
\onlinecite{horo-survey}. One of the most important topics in the theory
of entanglement is how to distinguish entangled states from separable
states. In the early eighties, Choi \cite{choi-ppt} observed that if
a positive (\lq positive\rq\ means \lq positive semi-definite\rq )
matrix in the tensor product belongs to tensor product of positive
parts then its partial transpose must be positive. Since the tensor
product of positive parts is just the convex cone generated by
separable states, this is equivalent to the PPT criterion which was
rediscovered later by Peres \cite{peres}. This criterion is already
implicit in the Woronowicz' earlier work \cite{woronowicz} who
initiated the duality theory between positivity of linear maps and
the separability of states. He constructed an example of positive partial transposed (PPT)
entangled state in order to show the existence of indecomposable
positive linear maps.

It was Horodecki's \cite{horo-1} who used the relation between positive maps and separable states to get
the complete criterion for separability. The Woronowicz's idea has been developed for general cases in Ref. \onlinecite{eom-kye}
to get the duality theory for $s$-positive linear maps. For the case of $s=1$, this is equivalent to the Horodecki's
criterion, through the Jamio\l kowski-Choi isomorphism \cite{jami,choi75-10}.
We note that the general cases give rise to the notion of Schmidt numbers and the relation to $s$-positivity of linear maps,
which was obtained independently in Ref. \onlinecite{t-h,sbl}. See also Refs.~\onlinecite{skowronek} and \onlinecite{stormer} for approaches to the duality theory which works for an arbitrary mapping cone.
Horodecki's criterion tells us that we need positive maps in order to detect entanglement, which has been reformulated in terms of entanglement witnesses \cite{terhal}. Under the JC isomorphism,  an entanglement
witness is just a positive linear map which is not completely positive.
An entanglement witness which detects a maximal set of entanglement is said to be optimal,
as was introduced in Ref. \onlinecite{lew00}.

In spite of its importance, the whole structure of the convex cone of all positive linear maps is far from
being completely understood, and there had even been very few known nontrivial examples of positive linear maps until the eighties.
To overcome this difficulty, the idea to consider the line segment from the
trace map to a given positive map had been used in mathematical literature to distinguish various notions of positivity. See Ref. \onlinecite{tom_83,tom_85}.
From the point of view of physics,
positive maps which are not completely positive, like the transpose maps, are non-physical operations. The main idea of structural physical approximation (SPA) \cite{horo_ekert,fiura,horo01} is to approximate positive maps by the nearest completely positive maps, which are physical operations, and so we can implement them  in the real world.
It was theoretically shown \cite{korbicz} and later demonstrated in practice \cite{lim_2,lim} that the SPA's of the transpose map and the partial transpose map are experimentally feasible. On the mathematics side, SPAs correspond to the points at which a segment between the trace map and a given positive map crosses the
border of the set of completely positive maps.

We recall that positive maps (respectively completely positive maps)
correspond to block-positive (respectively positive) matrices in
$M_m\otimes M_n$ under the JC isomorphism, where $M_n$ denotes the $*$-algebra of all
$n\times n$ matrices over the complex field, with the identity
$\openone_n$. We also note that the trace map corresponds to
$\openone_m \otimes\openone_n$. For an entanglement witness $W\in
M_m\otimes M_n$, we consider the line segment $L_W$ from $\openone_m\otimes\openone_n$ to $W$. The SPA of $W$ is the positive matrix on
$L_W$ nearest to $W$. It was conjectured in Ref. \onlinecite{korbicz} that if
$W$ is an optimal entanglement witness then its SPA is separable.
Several authors
\cite{chru_pyt_sra,chru_pyt_2,aug_bae,chru_pyt,cw,qi} considered
various classes of optimal entanglement witnesses to support the
conjecture. Here, we provide a counterexample.
Since non-decomposable optimal witnesses need not be optimal in the sense of Section IV of Ref.~\onlinecite{lew00}, we use the term PPTES witnesses \cite{ha_kye_opt_ind}. Here, PPTES refers to PPT entangled states.
PPTES optimality of a witness $W$ means that $W$ detects a maximal set (in the sense of inclusion) of PPT entangled states. This is equivalent to say that $W$ is
indecomposable and both $W$ and $W^\Gamma$ is optimal \cite{lew00}, where $W^{\Gamma}$ denotes the partial transpose of $W$. 
Let us recall that a special role is played by the product vectors $\phi\otimes\psi$ that satisfy
$\left<\phi\otimes\psi\right|W\left|\phi\otimes\psi\right>=0$. If these vectors span
$\mathbb C^m\otimes \mathbb C^n$, we say that $W$ has the \textit{spanning property}.
We say that $W$ is co-optimal (respectively co-spanning) if $W^\Gamma$ is
optimal (respectively spanning), bi-optimal if it is both optimal
and co-optimal, and bi-spanning similarly.

In order to deal with the SPA conjecture in a systematic way, we
introduce the notions of positive type and copositive type for
entanglement witnesses in the next section, and show that if the SPA
of an entanglement witness is separable, then the witness must be of
copositive type. We also see that $W$
is of positive type if and only if $W^\Gamma$ is of copositive type.
Because the optimality of PPTES witnesses is invariant under the
operation of partial transpose, we conclude that only approximately
half of optimal PPTES witnesses can satisfy the SPA conjecture. In the
third section, we exhibit examples to show that the SPA conjecture
fails even for the case of copositive type.
We consider entanglement witnesses given in Ref. \onlinecite{ha_kye_theta}
to find examples of optimal PPTES witnesses of
copositive type whose SPA are not separable. The SPA's of our
examples are exactly $3\otimes 3$ PPT edge states\cite{kye_osaka} of type $(6,8)$.
Very recently, after the authors posted this paper, St\o rmer \cite{stormer_spa}
also examined the same class of entanglement witnesses \cite{ha_kye_theta}
to find optimal indecomposable entanglement witnesses whose SPA are not separable.

\section{Positive and copositive type.}

From now on, we say that a block matrix is {\sl
copositive} if its partial transpose is positive. Copositive
matrices correspond to completely copositive maps under the
JC isomorphism. For a given block-positive matrix $W$, we consider
the line segment $L_W$ from $\openone_n\ot\openone_m$ to $W$, and
compare the distances to the nearest positive matrix and the nearest
copositive matrix on $L_W$. We say that $W$ is {\sl of positive}
(resp. {\sl copositive}) {\sl type} if the distance to the positive
(resp. copositive) part is shorter than or equal to the other. We
also say that a block-positive matrix is {\sl of PPT type} if two
distances coincide.

To be precise,
we consider self-adjoint matrix
$W_t=(1-t)/(mn)\openone_m\ot\openone_n+tW$.
When $t_0$ is the largest number in the interval $[0,1]$ for which $W_{t_0}$ is positive,
$W_{t_0}$ is said to be the SPA of $W$.
It is clear that the SPA of $W$ is separable if and only if the following condition
\begin{equation}\label{spa}
0\le t\le 1,\ W_t\ {\text{\rm is positive}}\
  \Longrightarrow \ W_t\ {\text{\rm is separable}}
\end{equation}
holds. It is also clear that the validity of the condition (\ref{spa}) is not changed when we replace
$1/(mn)\openone_m\otimes\openone_n$ in the definition of $W_t$ by $\openone_m\otimes\openone_n$ alone due to the convexity. Therefore, we may redefine
$W_t$ as
$$
W_t:=(1-t){\mathbb I_m\ot\mathbb I_n}+tW
$$
for every block-positive $W$,
 so far as we are concerned with the SPA conjecture.

Since separable states are of PPT, we see that if the SPA conjecture is true
then every optimal entanglement witness $W$ must satisfy the following condition
$$
0\le t\le 1,\ W_t\ {\text{\rm is positive}}\ \Longrightarrow \ W_t\ {\text{\rm is of PPT}}
$$
which is weaker than (\ref{spa}), and so satisfies the following equivalent condition
\begin{equation}\label{wspa}
0\le t\le 1,\ W_t\ {\text{\rm is positive}}\ \Longrightarrow \ W_t\ {\text{\rm is copositive}}.
\end{equation}
We define two real numbers $\alpha_W$ and $\beta_W$ in the interval $[0,1]$ by
$$
\begin{aligned}
\alpha_W:&=\sup\{ t\in [0,1]: W_t\ {\text{\rm is positive}}\},\\
\beta_W:&=\sup\{ t\in [0,1]: W_t\ {\text{\rm is copositive}}\},
\end{aligned}
$$
for an arbitrary block-positive $W$.
We see that both $\alpha_W$ and $\beta_W$ are nonzero, since $\mathbb I_m\ot\mathbb I_n$ is an interior point of the convex cone
generated by all separable states.
We also note that the number $1-\alpha_W$ (resp. $1-\beta_W$) plays a role of the distance from $W$ to the positive
(resp. copositive) part through the line segment $L_W$.
It is clear that $W$ satisfies the condition (\ref{wspa})  if and only if the inequality
$\alpha_W\le\beta_W$ holds.
Since $W_t$ is positive if and only if $W^\Gamma_t$ is copositive, we also have
$\alpha_W=\beta_{W^\Gamma}$ and $\alpha_{W^\Gamma}=\beta_W$.
Therefore, we see that $\alpha_W<\beta_W$ if and only if $\alpha_{W^\Gamma}>\beta_{W^\Gamma}$. In other words,
$W$ satisfies the condition (\ref{wspa}) if and only if $W^\Gamma$ does not satisfy (\ref{wspa}), whenever $\alpha_W\neq\beta_W$.

We see that a block-positive matrix $W$ is of positive type if and only if $\alpha_W\ge\beta_W$, and
of copositive type if and only if $\alpha_W\le\beta_W$ if and only if the condition (\ref{wspa}) holds.
We also see that $W$ is of PPT type if and only if $\alpha_W=\beta_W$.
The above discussion tells us that
$W$ is of positive type if and only if $W^\Gamma$ is of copositive type, and the SPA
conjecture can only hold for witnesses of copositive type. Especially, if the SPA of $W$ is separable, then $W$ is
of copositive type and the SPA of $W^\Gamma$ is never separable unless $W$ is of PPT type.
\begin{figure}[ht]
\includegraphics[scale=0.45]{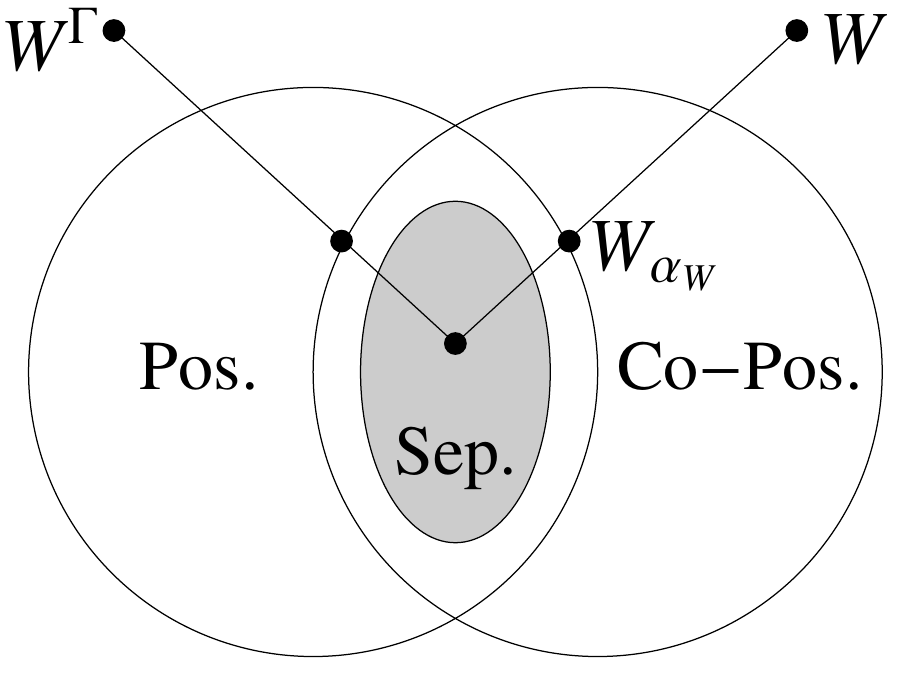}
\caption{$W$ has the mirror image $W^\Gamma$. $W$ is nearer to the copositive part if and only if $W^\Gamma$
is nearer to the positive part.}
\end{figure}

\section{Examples of PPT type.}

In this section, we exhibit examples of indecomposable entanglement witnesses of PPT type with the bi-spanning property in the sense
of Ref. \onlinecite{ha_kye_opt_ind}, whose SPAs are not separable.

For nonnegative real numbers $a,b,c$ and $-\pi\le\theta\le\pi$, we consider the following
self-adjoint block matrix in $M_3\ot M_3$:
$$
W[a,b,c;\theta]=
\left(
\begin{array}{ccccccccccc}
a     &\cdot   &\cdot  &\cdot  &-e^{i\theta}     &\cdot   &\cdot   &\cdot  &-e^{-i\theta}     \\
\cdot   &c &\cdot    &\cdot    &\cdot   &\cdot &\cdot &\cdot     &\cdot   \\
\cdot  &\cdot    &b &\cdot &\cdot  &\cdot    &\cdot    &\cdot &\cdot  \\
\cdot  &\cdot    &\cdot &b &\cdot  &\cdot    &\cdot    &\cdot &\cdot  \\
-e^{-i\theta}     &\cdot   &\cdot  &\cdot  &a     &\cdot   &\cdot   &\cdot  &-e^{i\theta}     \\
\cdot   &\cdot &\cdot    &\cdot    &\cdot   &c &\cdot &\cdot    &\cdot   \\
\cdot   &\cdot &\cdot    &\cdot    &\cdot   &\cdot &c &\cdot    &\cdot   \\
\cdot  &\cdot    &\cdot &\cdot &\cdot  &\cdot    &\cdot    &b &\cdot  \\
-e^{i\theta}     &\cdot   &\cdot  &\cdot  &-e^{-i\theta}     &\cdot   &\cdot &\cdot  &a
\end{array}
\right).
$$
We also put
$p_\theta=\max\{ q_{(\theta-\frac 23 \pi)}, q_\theta, q_{(\theta+\frac 23 \pi)}\}$,
where $q_\theta=e^{i\theta}+e^{-i\theta}$. We note that $1\le p_\theta\le 2$. We also see that
$p_\theta=2$ if and only if $\theta=0,\,\pm 2\pi/3$,
and $p_\theta=1$ if and only if $\theta=\pm\pi/ 3,\,\pm\pi$.
It is easy to see that $W[a,b,c;\theta]$ is of PPT if and only if
\begin{equation}\label{ppt}
a\ge p_\theta,\qquad bc\ge 1.
\end{equation}
If $1< p_\theta<2$,
the cases of $a=p_\theta$ and $bc=1$ give rise to new kinds of examples \cite{kye_osaka}
of PPT edge states of type $(8,6)$. 
If $\theta=0$ then the notion of PPT coincides \cite{kye_osaka} with  separability.
On the other hand, the authors \cite{ha_kye_geom_sep}
explored the boundary structures between the separability and the inseparability for PPT states
when $\theta=\pi$. We recall that the case $\theta=0$ had been considered in Ref. \onlinecite{cho-kye-lee}.

It turns out \cite{ha_kye_theta} that $W[a,b,c;\theta]$ is block-positive if and only if
the condition
\begin{equation}\label{positive}
a+b+c\ge p_\theta,\qquad a\le 1\Longrightarrow bc\ge (1-a)^2
\end{equation}
holds.
We refer to Ref. \onlinecite{ha_kye_theta} for the pictures of the $3$-dimensional convex bodies determined
by (\ref{ppt}) and (\ref{positive}) for fixed $\theta$.
From now on, we concentrate on the case
$a<p_\theta$ and $bc<1$,
for which $W[a,b,c;\theta]$ is neither positive nor copositive.
We have
$$
W_t[a,b,c;\theta]=
t \, W\left[\dfrac {a_t}t,\dfrac{b_t}t,\dfrac{c_t}t;\theta\right],\qquad 0<t\le 1,
$$
with
$a_t=1-t+ta$, $b_t=1-t+tb$, and $c_t=1-t+tc$.
We see that $W_t[a,b,c;\theta]$ is positive
if and only if $a_t\ge tp_\theta$ if and
only if $t\le \alpha_W=1/(p_\theta+1-a)$,
and $W_t[a,b,c;\theta]$ is copositive if and only if $b_tc_t\ge t^2$ if and only if
$$
F(t):=(b+c-bc)t^2-(b+c-2)t-1\le 0.
$$
Therefore, we see that $W[a,b,c;\theta]$ is of copositive type if and only if
$0\le t\le \alpha_W$ implies $F(t)\le 0$.
Since $F(0)=-1<0$ and $F(1)=1-bc>0$, we see that this happens if and only if
$F(\alpha_W)\le 0$ if and only if
$$
(p_\theta-a+b)(p_\theta-a+c)\ge 1.
$$
Analogously, $W[a,b,c;\theta]$ is of positive type if and only if the reverse inequality holds,
and of PPT type if and only if the equality holds.

On the other hand, we note that the SPA of $W[a,b,c;\theta]$ is given by
$$
W_{\alpha_W}[a,b,c;\theta]
=\dfrac1{p_\theta+1-a}W[p_\theta, p_\theta-a+b,p_\theta-a+c;\theta].
$$
We note \cite{kye_osaka} that $W_{\alpha_W}[a,b,c;\theta]$
is not separable when
\begin{equation}\label{pptes}
(p_\theta-a+b)(p_\theta-a+c)=1,\qquad 1< p_\theta<2,
\end{equation}
that is when $W[a,b,c;\theta]$ is of PPT type.
We concentrate on the following two subcases:
\begin{eqnarray}
\label{case3} 2-p_\theta\le a<1,\quad a+b+c= p_\theta,\quad bc= (1-a)^2,\,\\
\label{case1} 1\le a< p_\theta,\quad a+b+c= p_\theta,\quad bc=0.\quad \quad \quad \
\end{eqnarray}
We note \cite{ha_kye_theta}
that $W[a,b,c;\theta]$ is indecomposable in the above cases;
has the bi-spanning property in the case (\ref{case3})
and has the co-spanning property in the  case (\ref{case1}).
See Ref. \onlinecite{ha+kye_indec-witness,ha+kye_exposed} for the case of $\theta=0$.
We also note that condition \eqref{pptes} together with \eqref{case3} is translated into
\begin{equation*}
3a^2-2(2p_\theta+1)a+2p_\theta^2=0,\quad  2-p_\theta\le a<1,
\end{equation*}
and it is easy to see that
\begin{enumerate}
\item[(i)]
There exists $a,b$ and $c$ satisfying (\ref{pptes}) and (\ref{case3}) if and only if $4/3\le p_{\theta}<1+1/{\sqrt2}$.
\end{enumerate}
In this case, there are two solutions as is seen in FIG 2 when $2-p_\theta< a<1$. If $a=2-p_\theta$ then we have only one solution
with $b=c=p_\theta-1$.
For the case \eqref{case1}, we note that the condition \eqref{pptes} is equivalent to
\begin{equation*}
2(p_{\theta}-a)^2=1,\quad 1\le a < p_{\theta}<2,
\end{equation*}
and thus we have the following:
\begin{enumerate}
\item[(ii)]
There exists $a,b$ and $c$ satisfying (\ref{pptes}) and (\ref{case1}) if and only if $1+1/{\sqrt2}\le p_\theta<2$.
\end{enumerate}
\begin{figure}[ht]
\includegraphics[scale=0.36]{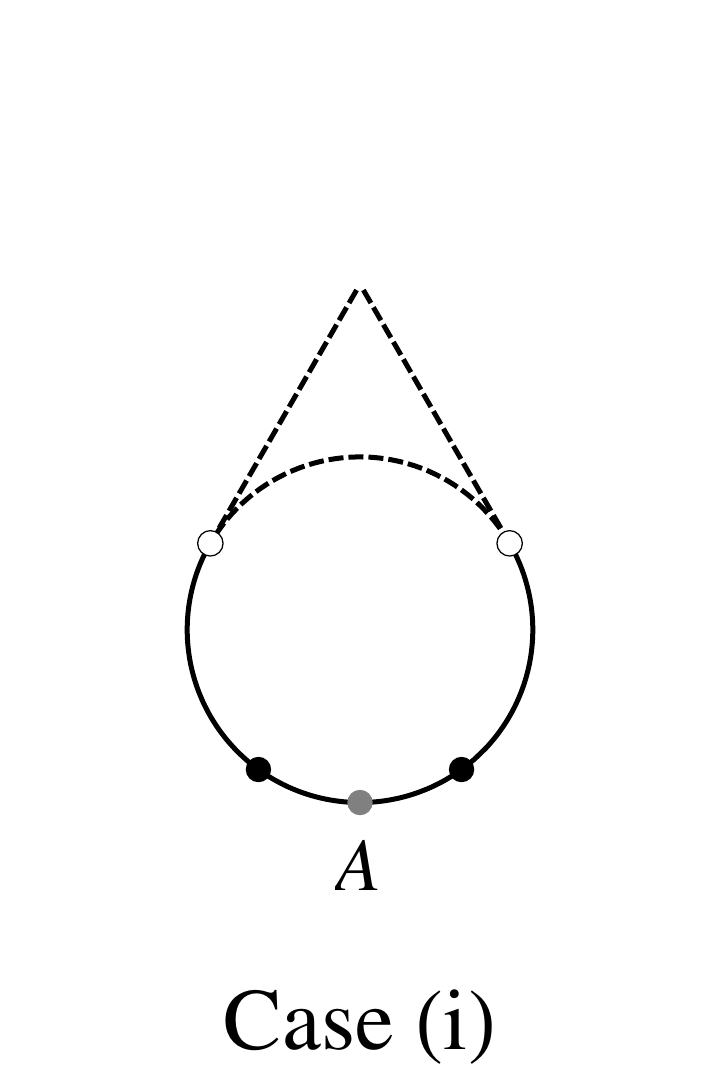} \quad \quad
\includegraphics[scale=0.36]{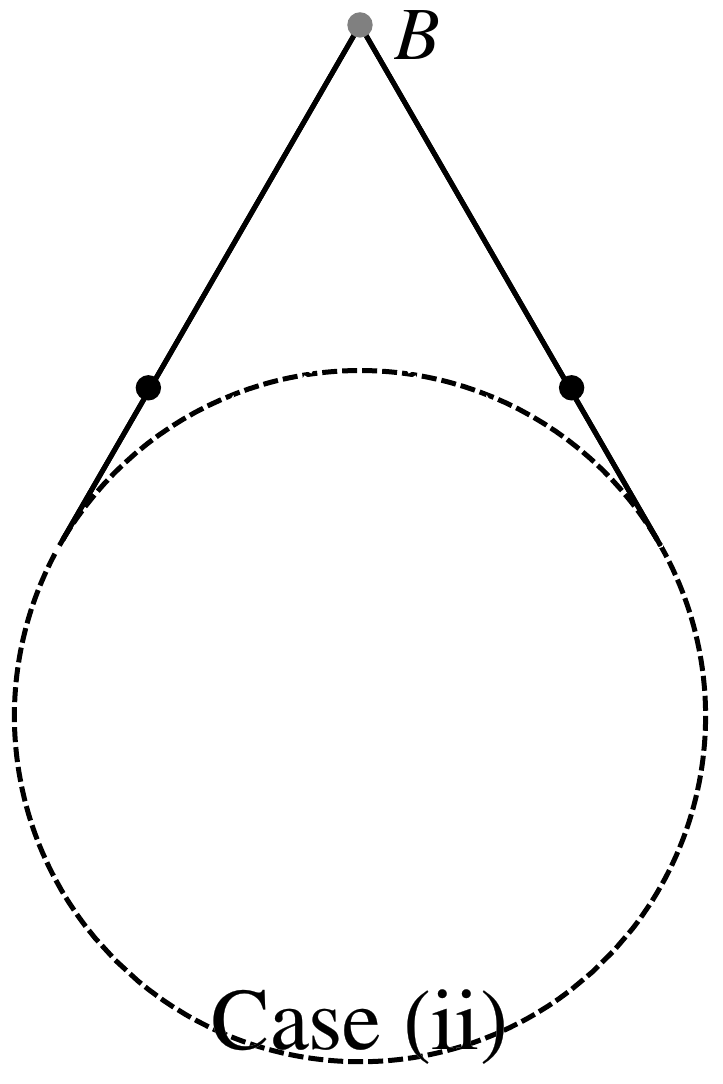}
\caption{Figure of the boundary of the convex body determined by  (\ref{ppt}) and (\ref{positive}), which is  lying on the plane $a+b+c=p_{\theta}$ for fixed $\theta$.
Continous line in the case (i)
(respectively (ii)) indicates $a,\,b,\,c$ satisfying the condition~\eqref{case3}
(respectively \eqref{case1}). In both cases, black dots represent PPT types which violate the SPA conjecture. We refer to FIG. 3 of  Ref.~\onlinecite{ha_kye_theta} for the pictures of the $3$-dimensional convex bodies.
The point $A$ represents the case $a=2-p_\theta, b=c=p_\theta-1$, and the point $B$ represents
the case $a=p_\theta, b=c=0$. We refer to the figure in Ref. \onlinecite{kye_ritsu}, Section 5 for the case of $\theta=0$.}
\end{figure}

In the first case (i), we have indecomposable entanglement witnesses with the bi-spanning properties whose SPA are not separable.
These entanglement witnesses are optimal PPTES witnesses in the
sense of Ref. \onlinecite{ha_kye_opt_ind} which detect maximal sets of PPTES.
In the case (ii), we see that the partial transposes $W^\Gamma$ give
rise to examples of optimal indecomposable entanglement witnesses
whose SPA are not separable. It should be noted that they are not
\lq nd-OWE\rq s in the sense of Ref. \onlinecite{lew00}, or PPTES optimal witnesses using our terminology, when $1+1/{\sqrt2}<
p_\theta<2$, because they are not co-optimal; the smallest face
determined by $W^\Gamma$ contains $W[p_\theta,0,0;\theta]^\Gamma$
which is copositive. If $p_\theta=1+1/{\sqrt 2}$ then we have the solution
$a=1, b=p_\theta-1, c=0$ or $a=1,b=0,c=p_\theta-1$, for which
$W[a,b,c;\theta]$ are  bi-optimal entanglement witnesses without the spanning property \cite{ha_kye_theta}.

\section{Conclusion.}

In conclusion,
we propose the notions of positive type and copositive type for block-positive matrices $W$. We have shown that
if the SPA of $W$ is separable then $W$ must be of copositive type, and $W$ is copositive type if and only if its partial transpose $W^\Gamma$
is of positive type. Furthermore, a PPTES witness $W$ is optimal if and only if $W^\Gamma$ is optimal. Therefore, using the above facts about positive and copositive types, we conclude that the separability of the SPA of an optimal PPTES witness $W$ implies that the SPA of $W^{\Gamma}$ is not separable, unless $W$ is of PPT type. This is so, even though $W^{\Gamma}$ is still an optimal PPTES witness.
We provide concrete examples of indecomposable entanglement witnesses of both copositive type and PPT type which violate the SPA conjecture.
It would be very nice to characterize entanglement witnesses whose SPA are separable.

Especially, it is still open if the SPA of an optimal decomposable entanglement witness is separable or not.
Recent development \cite{kye_dec_wit,asl}
on the optimality for decomposable cases might be helpful to determine if the SPA conjecture is true for decomposable case.

\begin{acknowledgments}
KCH is partially supported by NRFK 2012-0002600. SHK is partially supported by NRFK 2012-0000939.
This is a revised and expanded version of the preprint circulated under the same title, authored by SHK alone.
Examples of PPT type are included in this revised version. The authors are grateful to
Antonio Acin,
Mafalda Almeida,
Remigiusz Augusiak,
Joonwoo Bae,
Jaroslaw Korbicz
and
Maciej Lewenstein
for their feedback on the preprint. Special thanks are due to
Joonwoo Bae who informed us their discussion about the preprint, and
Remigiusz Augusiak who pointed out errors in the preprint.
The authors are also grateful to Hyang-Tag Lim for helpful discussion on the papers \cite{lim_2,lim}.
\end{acknowledgments}


\begin{thebibliography}{999}

\bibitem{aug_bae}
R. Augusiak, J. Bae, \L . Czekaj, and M. Lewenstein, ``On structural
physical approximations and entanglement breaking maps,''
 J. Phys. A {\bf 44}, 185308 (2011).

\bibitem{asl}
R. Augusiak, G. Sarbicki, and M. Lewenstein, ``Optimal decomposable
witnesses without the spanning property,'' Phys. Rev. A {\bf 84},
052323 (2011).

\bibitem{cho-kye-lee}
S.-J. Cho, S.-H. Kye, and S. G. Lee, ``Generalized Choi maps in
three-dimensional matrix algebra,'' Linear Algebra Its Appl. {\bf
171}, 213--224 (1992).

\bibitem{choi75-10}
M.-D. Choi, ``Completely positive linear maps on complex matrices,''
Linear Algebra Its Appl. {\bf 10}, 285--290 (1975).

\bibitem{choi-ppt}
M.-D. Choi, {\it Operator Algebras and Applications} (Kingston,
1980), pp. 583--590, Proc. Sympos. Pure Math. Vol 38. Part 2, Amer.
Math. Soc., 1982.

\bibitem{chru_pyt_2}
D. Chru\'{s}ci\'{n}ski, and J. Pytel, ``Constructing optimal
entanglement witnesses. II. Witnessing entanglement in 4N×4N
systems,'' Phys. Rev. A {\bf 82}, 052310 (2010).

\bibitem{chru_pyt}
D. Chru\'{s}ci\'{n}ski and J. Pytel, ``Optimal entanglement
witnesses from generalized reduction and Robertson maps,'' J. Phys.
A {\bf 44}, 165304 (2011).

\bibitem{chru_pyt_sra}
D. Chru\'{s}ci\'{n}ski, J. Pytel, and G. Sarbicki, ``Constructing
optimal entanglement witnesses,'' Phys. Rev. A {\bf 80}, 062314
(2009), 062314.

\bibitem{cw}
D. Chru\'{s}ci\'{n}ski and F. A. Wudarski, ``Geometry of
Entanglement Witnesses for Two Qutrits,'' Open Syst. Inf. Dyn. {\bf
18}, 375--387 (2011).

\bibitem{eom-kye}
M.-H. Eom, and S.-H. Kye, ``Duality for positive linear maps in
matrix algebras,'' Math. Scand. {\bf 86}, 130--142 (2000).

\bibitem{fiura}
J. Fiurasek, ``Structural physical approximations of unphysical maps
and generalized quantum measurements,'' Phys. Rev. A {\bf 66},
052315 (2002).

\bibitem{ha+kye_indec-witness}
K.-C. Ha and S.-H. Kye, ``One-parameter family of indecomposable
optimal entanglement witnesses arising from generalized Choi maps,''
Phys. Rev. A {\bf 84}, 024302 (2011).

\bibitem{ha+kye_exposed}
K.-C. Ha and S.-H. Kye, ``Entanglement Witnesses Arising from
Exposed Positive Linear Maps,'' Open Syst. Inf. Dyn. {\bf 18},
323--337 (2011).

\bibitem{ha_kye_geom_sep}
K.-C. Ha and S.-H. Kye, ``Geometry of the faces for separable states
arising from generalized Choi maps,'' Open Syst. Inf. Dyn. {\bf 19},
125009 (2012).

\bibitem{ha_kye_opt_ind}
K.-C. Ha and S.-H. Kye, ``Optimality for indecomposable entanglement
witnesses,'' Phys. Rev. A (to appear), e-print arXiv:1204.6596. 

\bibitem{ha_kye_theta}
K.-C. Ha and S.-H. Kye, ``Entanglement witnesses arising from Choi
type positive linear maps,'' e-print arXiv:1205.2921.

\bibitem{horo-1}
M. Horodecki, P. Horodecki, and R. Horodecki, ``Separability of
mixed states: necessary and sufficient conditions,'' Phys. Lett. A
{\bf 223}, 1--8 (1996).

\bibitem{horo01}
P. Horodecki, ``From limits of quantum operations to multicopy
entanglement witnesses and state-spectrum estimation,'' Phys. Rev. A
{\bf 68}, 052101 (2003).

\bibitem{horo_ekert}
P. Horodecki, and A. Ekert, ``Method for Direct Detection of Quantum
Entanglement,'' Phys. Rev. Lett. {\bf 89}, 127902 (2002).

\bibitem{horo-survey}
R. Horodecki, P. Horodecki, M. Horodecki, and K. Horodecki, ``Quantum entanglement,''
Rev. Mod. Phys. {\bf 81}, 865--942 (2009).

\bibitem{jami}
A. Jamio\l kowski, ``An effective method of investigation of
positive maps on the set of positive definite operators,'' Rep.
Math. Phys. {\bf 5}, 415--424 (1974).

\bibitem{korbicz}
J. K. Korbicz, M. L. Almeida, J. Bae, M. Lewenstein, and A. Acin,
``Structural approximations to positive maps and
entanglement-breaking channels,'' Phys. Rev. A {\bf 78}, 062105
(2008), 062105.

\bibitem{kye_ritsu}
S.-H. Kye, ``Facial structures for various notions of positivity and
applications to the theory of entanglement,'' e-print
arXiv:1202.4255.

\bibitem{kye_dec_wit}
S.-H. Kye, ``Necessary conditions for optimality of decomposable
entanglement witness,'' Rep. Math. Phys(to be published), e-print
arXiv:1108.0456.

\bibitem{kye_osaka}
S.-H. Kye and H. Osaka, ``Classification of bi-qutrit positive
partial transpose entangled edge states by their ranks,'' J. Math.
Phys. {\bf 53}, 052201 (2012)

\bibitem{lew00}
M. Lewenstein, B. Kraus, J. I. Cirac, and P. Horodecki,
``Optimization of entanglement witnesses,'' Phys. Rev. A {\bf 62},
052310 (2000).

\bibitem{lim_2}
H.-T. Lim, Y.-S. Kim, Y.-S. Ra, J. Bae and Y.-H. Kim, ``Experimental
Realization of an Approximate Partial Transpose for Photonic
Two-Qubit Systems,'' Phys. Rev. Lett. {\bf 107}, 160401 (2011).

\bibitem{lim}
H.-T. Lim, Y.-S. Ra, Y.-S. Kim, J. Bae, and Y.-H. Kim,
``Experimental implementation of the universal transpose operation
using the structural physical approximation,'' Phys. Rev. A {\bf
83}, 020301 (2011).

\bibitem{peres}
A. Peres, ``Separability Criterion for Density Matrices,''
 Phys. Rev. Lett. {\bf 77}, 1413 (1996).

\bibitem{qi}
X. Qi, and J. Hou, ``Characterization of optimal entanglement
witnesses,'' Phys. Rev. A {\bf 85}, 022334 (2012).

\bibitem{sbl}
A. Sanpera, D. Bru\ss, and M. Lewenstein, ``Schmidt-number witnesses
and bound entanglement,'' Phys. Rev. A {\bf 63}, 050301 (2001).

\bibitem{skowronek} L. Skowronek, ``Cones with a mapping cone symmetry in the finite-dimensional case,'' Linear Algebra Its Appl. {\bf 435}, 361--370 (2011).

\bibitem{stormer} E. St\o rmer, ``Duality of cones of positive maps,'' M\"unster J. Math. {\bf 2}, 299--310 (2009).

\bibitem{stormer_spa}
E. St\o rmer,
``Separable states and SPA of a positive map,''
e-print arXiv:1206.563.


\bibitem{terhal}
B. M. Terhal, ``Bell inequalities and the separability criterion,''
Phys. Lett. A {\bf 271}, 319--326 (2000).

\bibitem{t-h}
B. M. Terhal, and P. Horodecki, ``Schmidt number for density
matrices,'' Phys. Rev. A {\bf 61}, 040301 (2000).

\bibitem{tom_83}
T. Takasaki and J. Tomiyama, ``On the geometry of positive maps in
matrix algebras,'' Math. Z. {\bf 184}, 101--108 (1983).

\bibitem{tom_85}
J. Tomiyama, ``On the geometry of positive maps in matrix algebras.
II,'' Linear Algebra Its Appl. {\bf 69}, 169--177 (1985).

\bibitem{tom_86}
J. Tomiyama, ``On the geometry of positive maps in matrix
algebras,'' Contemp. Math. {\bf 62}, 357 (1987).

\bibitem{woronowicz}
S. L. Woronowicz, ``Positive maps of low dimensional matrix algebras,''
Rep. Math. Phys. {\bf 10}, 165--183 (1976).


























\end{thebibliography}
\end{document}